\documentclass[conference]{IEEEtran}
\IEEEoverridecommandlockouts
\usepackage{cite}
\usepackage{amsmath,amssymb,amsfonts}
\usepackage{graphicx}
\usepackage{textcomp}
\usepackage{booktabs}
\usepackage[ruled,vlined]{algorithm2e}
\usepackage{balance}
\usepackage{url}
\usepackage{microtype}
\newcommand{\Det}{Det-5G}
\begin{document}

\title{Det-5G: Closing the Determinism Gap in 5G-Advanced for Industrial Closed-Loop Control}

\author{\IEEEauthorblockN{Adnan Aijaz}
\IEEEauthorblockA{Bristol Research and Innovation Laboratory, Toshiba Europe Ltd., Bristol, United Kingdom\\
adnan.aijaz@toshiba-bril.com}}

\maketitle

\begin{abstract}
The ultra-reliable low-latency communication (uRLLC) capability of 5G has created significant opportunities for industrial wireless connectivity, yet widespread use of cellular networks for closed-loop control remains challenging. Closed-loop control requires more than low packet latency and high reliability: cyclic command/feedback exchanges must complete within predictable time bounds despite changing channel conditions, recovery transmissions, mobility, and multi-device contention. This paper introduces Deterministic-5G (\Det), a unified radio resource allocation framework for industrial closed-loop control. \Det{} treats the complete bidirectional control cycle as the scheduling object and combines coordinated downlink/uplink allocation, adaptive bundled transmissions, group-oriented downlink communication, and optimized multi-user uplink scheduling over 5G air-interface. Its performance is evaluated through a combination of closed-form analysis and Monte Carlo scheduling experiments, with comparisons against conventional dynamic grant-based scheduling, semi-persistent scheduling/configured grant operation, and fixed proactive repetition. The evaluation shows that \Det{} improves predictability of cycle completion, maintains the target reliability under changing link conditions, and scales more effectively to multi-device control than conventional reactive scheduling, while adapting radio resource use instead of continuously provisioning for the worst case as in fixed repetition. These characteristics make cycle-oriented scheduling a pragmatic solution for reducing the determinism gap that limits the use of 5G for closed-loop control in different verticals, especially as it evolves through 5G-Advanced toward 6G.
\end{abstract}

\begin{IEEEkeywords}
5G, 6G, closed-loop control, determinism, Industry 5.0,  private networks, resource allocation, scheduling, uRLLC.
\end{IEEEkeywords}

\section{Introduction}
5G has been positioned as a key wireless technology for industrial automation because uRLLC, private/non-public networks, and time-sensitive communication extend cellular connectivity toward applications traditionally served by deterministic wired systems \cite{pvt_5G,3gpp_tsn}. Yet low packet latency and high reliability are not sufficient to make a wireless system deterministic for control. A controller repeatedly issues a command and expects sensing or actuation feedback within the same control cycle. The relevant performance object is therefore completion of a causally linked closed-loop transaction, not latency of an isolated packet. Independent downlink/uplink scheduling and reactive retransmissions can make cycle completion unpredictable even when average packet latency is low.

This gap matters for industrial adoption. Stringent motion-control profiles combine millisecond-level transfer intervals and latency with essentially no tolerance for missed cycles \cite{3gpp_22104,5gacia_sls}, while industrial guidance identifies non-deterministic timing as a production risk \cite{5gacia_test}. The issue is timely in 5G-Advanced: private 5G has enabled industrial connectivity, but adoption for production-critical control remains hindered by unpredictable command/feedback completion.

This paper presents Deterministic-5G (\Det{}), a radio resource allocation solution that revisits industrial 5G scheduling with completion of the closed-loop control cycle as the primary objective. Its core design principle is that a command and its corresponding feedback form a single causally coupled transaction whose timing and reliability should be considered jointly, rather than as independent downlink and uplink deliveries. The paper makes the following contributions:
\begin{itemize}
\setlength{\itemsep}{1pt}
\setlength{\parskip}{0pt}
\setlength{\parsep}{0pt}

\item \emph{Cycle-level resource allocation and reliability control:} a self-contained cyclic transmission procedure jointly provisions resources for command delivery, feedback, proactive redundancy, acknowledgement, and recovery. It supports both periodically reserved and cycle-by-cycle operation to balance signaling overhead and adaptability.

\item \emph{Single- and multi-device industrial control:} the cycle-level design supports both individual control loops, representative of conventional industrial machines and production equipment, and coordinated groups of devices required by emerging applications such as wireless robot control, collaborative robotics, and distributed production systems. For multi-device operation, the framework incorporates common downlink delivery, selective recovery, and optimized sharing of uplink time-frequency resources.

\item \emph{Determinism-oriented performance evaluation:} analytical and Monte Carlo methods evaluate complete-cycle behavior against dynamic grant-based scheduling, SPS/CG, and fixed proactive repetition, considering reliability, recovery-induced timing variation, multi-device scalability, air-interface granularity, and mobility.
\end{itemize}


Det-5G uses existing New Radio (NR) mechanisms rather than a new 5G-Advanced primitive. As 5G-Advanced evolves NR toward 6G \cite{3gpp_rel20}, addressing the determinism gap within the existing air interface becomes increasingly important for robotics and Physical AI \cite{li_networked_robotics_6g}.

\section{Related Work}

Early work on industrial closed-loop control focused largely on non-cellular wireless systems. ENCLOSE introduced an enhanced single-hop interface for factory automation, while GALLOP targeted high-performance wireless closed-loop control for industrial IoT \cite{enclose,gallop}. These studies showed the importance of designing communication around periodic command/feedback exchange, tight timing, and reliability, but did not address cellular mechanisms such as 5G grant procedures or NR resource allocation.

Within 5G, uRLLC research has extensively studied latency, reliability, retransmission, and scheduling. Prior work has examined factory-automation requirements \cite{5G_FA}, retransmission-aware allocation and communication-control co-design \cite{retrans_URLLC,URLLC_control1}, scheduler-induced burst errors for periodic traffic \cite{demel2020burst}, and semi-persistent/configured-grant operation for time-sensitive traffic \cite{sch_TSN1,zhang2023cg}.

More general industrial schedulers provide per-flow guarantees or joint uplink/downlink allocation \cite{zhang2023rtss,kleinberger2026flex}, but typically retain packet- or flow-level scheduling with reactive recovery. The gap addressed here is the lack of a unified cycle-oriented treatment that combines proactive reliability, cycle-level recovery, group downlink delivery, and scalable multi-device resource allocation within 5G NR.

\section{5G NR and Resource Allocation Preliminaries}
\subsection{NR time structure}
\begin{figure}[t]
\centering
\includegraphics[width=0.96\columnwidth]{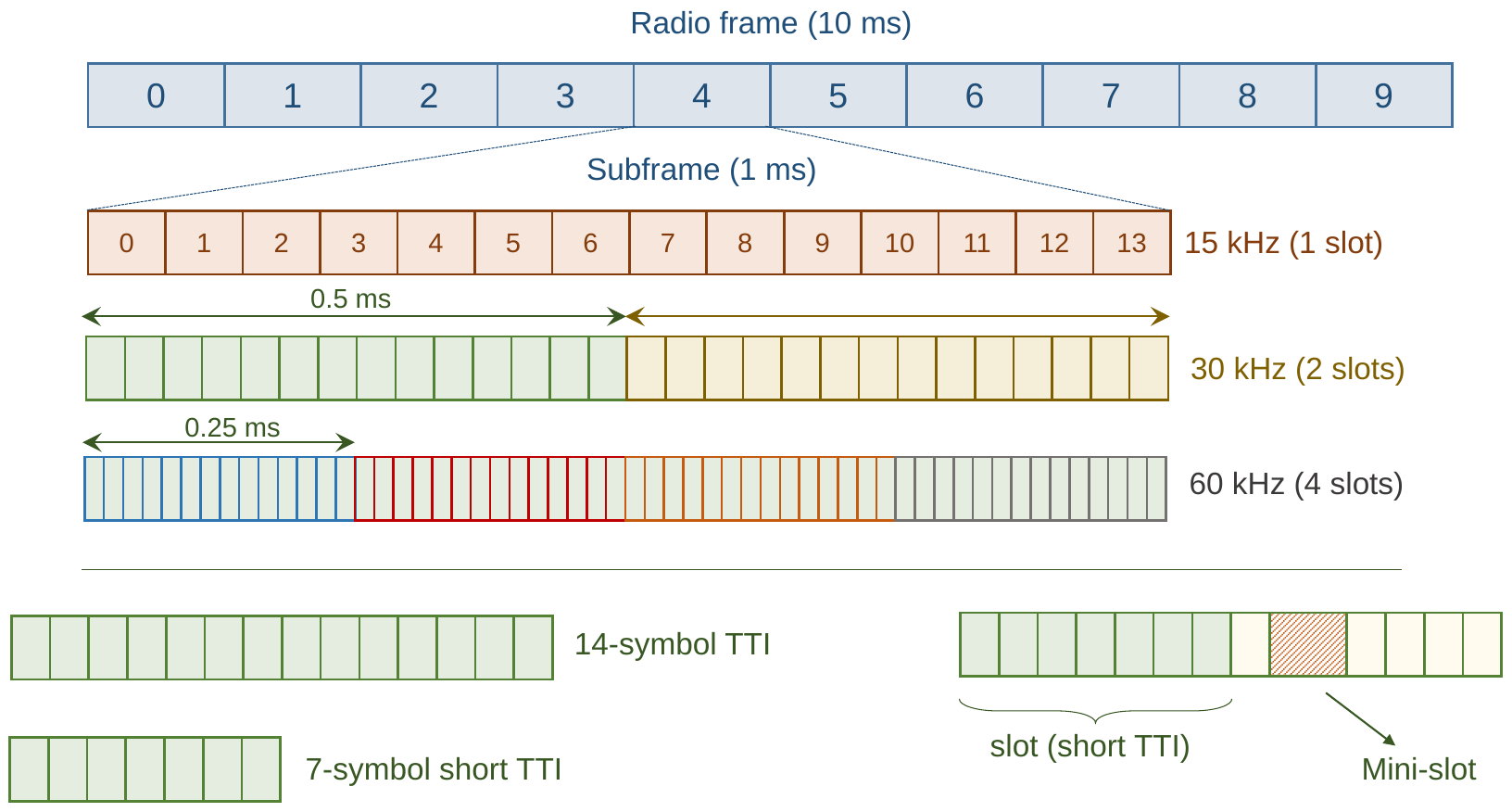}
\caption{Illustration of scalable NR numerology and SBT/NSBT opportunities.}
\label{slots}
\end{figure}

5G NR supports scalable OFDM numerology with SCS $\Delta f=15\times2^{\mu}$~kHz. With normal cyclic prefix, a slot contains 14 OFDM symbols and its duration decreases as SCS increases. NR also permits shorter transmissions over fewer symbols, including 2-, 4-, and 7-symbol transmission intervals, allowing non-slot-based placement at finer time granularity \cite{3gpp.38.214}. We use \emph{slot-based transmission} (SBT) for allocation over a complete selected transmission interval and \emph{non-slot-based transmission} (NSBT) for flexible symbol-level placement. Fig.~\ref{slots} illustrates these opportunities.

\subsection{Conventional scheduling for cyclic control}
In conventional dynamic operation, downlink and uplink are scheduled independently; uplink additionally requires scheduling request/grant signaling before data transmission. HARQ improves reliability reactively through decoding feedback and retransmission. For periodic traffic, semi-persistent scheduling (SPS) can reserve recurring downlink resources and configured grant (CG) scheduling can preconfigure uplink resources; we refer to their combined periodic operation as SPS/CG. These mechanisms reduce recurring control overhead, but they do not intrinsically bind the two directions into one command/feedback transaction. When HARQ is invoked, cycle completion depends on random decoding outcomes. Fixed proactive repetition removes this reactive timing uncertainty, but a small repetition factor becomes unreliable as the channel degrades whereas a worst-case factor consumes the same resources in every cycle. \Det{} is designed to jointly address cycle predictability, target reliability, and adaptive resource use.

\section{Det-5G: Design and Protocol Operation}
\subsection{System model and design principles}
\begin{figure}[t]
\centering
\includegraphics[width=0.94\columnwidth]{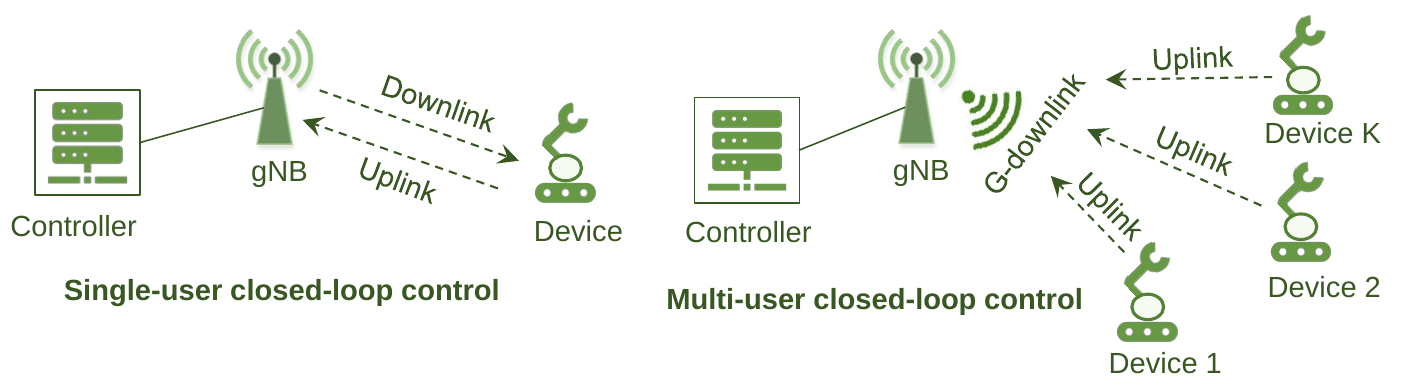}
\caption{Single-user and multi-user closed-loop control scenarios for \Det{}.}
\label{sys_model}
\end{figure}

Fig.~\ref{sys_model} shows the two scenarios considered by \Det{}: single-user closed-loop control and multi-user closed-loop control. The controller is connected to the gNB through a local wired/edge path, while the controlled device(s) use the NR air interface. The controller may execute at an industrial edge platform or elsewhere in the local network; the scope here is the radio-interface part of the control cycle. Devices are assumed to be connected before cyclic operation starts, since connection setup does not belong to the recurring control deadline.

\Det{} follows four design principles. First, \emph{downlink/uplink co-design} recognizes that command and feedback are causally coupled and should not be scheduled as unrelated packets. Second, \emph{self-contained cyclic allocation} reserves the control and data resources needed to complete one cycle. Third, \emph{bundled low-latency transmissions} place reliability resources before the deadline, with a bundle length that adapts to link conditions. Fourth, \emph{multi-user optimization} uses group communication in downlink and time-frequency packing in uplink so that cycle time does not grow linearly with the number of controlled devices.

\subsection{Single-user closed-loop control}
\begin{figure}[t]
\centering
\includegraphics[width=0.94\columnwidth]{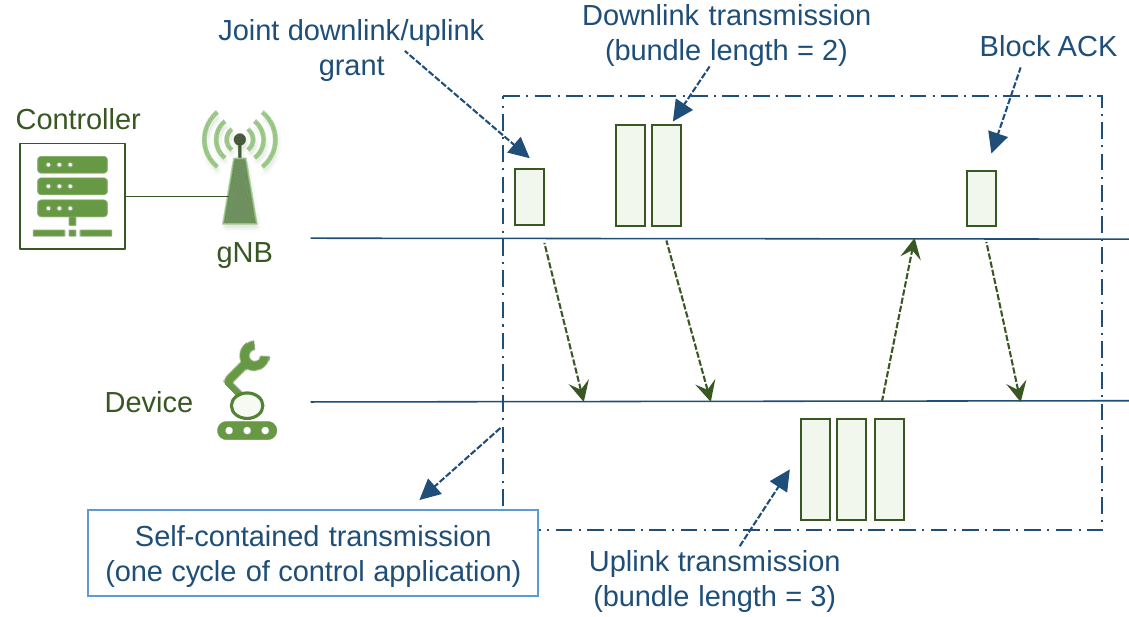}
\caption{Self-contained cyclic transmission for single-user \Det{}.}
\label{su_self_cont}
\end{figure}

A typical cycle consists of a controller command followed by sensing or actuation feedback from the device; the same framework applies if the application reverses the direction order. The fundamental \Det{} scheduling object is the \emph{self-contained transmission} in Fig.~\ref{su_self_cont}. It begins with a joint downlink/uplink resource allocation that identifies resources for the downlink command, the subsequent uplink response, and the final cycle-level acknowledgement/control. The downlink may contain one transmission or a bundle of repeated transmissions. After the downlink phase, the device transmits its uplink response, again as a single transmission or a bundle. A block acknowledgement is issued after the uplink bundle rather than after each individual element. Thus, the resources required for one complete control transaction are determined together.

A self-contained cycle can be realized entirely with SBTs or with a combination of SBTs and NSBTs. Once cyclic control starts, the allocation can operate in two ways. In \emph{reserved operation}, the downlink/uplink resources of the self-contained transmission repeat periodically, eliminating recurring grant exchange for stable traffic. In \emph{cycle-by-cycle operation}, a joint allocation is provided for the next cycle or group of cycles. The latter allows bundle length and resource placement to track mobility, interference, or changing link conditions. This distinction is important: periodic reservation reduces signaling, whereas cycle-by-cycle operation maximizes adaptability without abandoning the self-contained-cycle structure.

The complete cycle may itself be repeated when the transaction fails. Failure occurs if the downlink command is not decoded and therefore no valid uplink response is generated, or if the uplink bundle cannot be decoded. Instead of sending an ordinary block acknowledgement, the gNB can issue a new joint allocation for repeating the self-contained cycle. Bundling is also applicable to control information when additional robustness is required.

\Det{} further uses bundled transmissions as a \emph{proactive HARQ} mechanism. Reliability resources are placed before waiting for a conventional HARQ feedback/recovery round. A bundle may repeat coded data, may distribute data and additional FEC across its elements, or may use elements with different FEC contributions for the same payload. The key property is not a particular coding realization; it is that the scheduler knows the reliability expenditure and the corresponding cycle budget before transmission begins.

\subsection{Single-user bundle-length adaptation}
A large bundle improves reliability but consumes time-frequency resources. \Det{} therefore adapts downlink and uplink bundle lengths using recent received-signal statistics. Let $\mathrm{RSSI}_{\rm Avg}$ and $\mathrm{RSSI}_{\rm Var}$ denote the mean and variance observed for the most recent bundle, $\mathrm{RSSI}_{K}$ the moving average over the previous $K$ bundles, and $\mathrm{RSSI}_{\rm Th}$ and $\mathrm{RSSI}_{\rm VarTh}$ design thresholds. An update is triggered when
\begin{equation}
\begin{aligned}
&\mathrm{RSSI}_{\rm Avg}<\mathrm{RSSI}_{\rm Th}\ \lor\
\mathrm{RSSI}_{\rm Avg}<\mathrm{RSSI}_{K}\ \lor\\
&\mathrm{RSSI}_{\rm Var}>\mathrm{RSSI}_{\rm VarTh}.
\end{aligned}
\label{eq:update}
\end{equation}
The closed-loop exchange provides fresh measurements continuously; for example, downlink quality information can be piggybacked on the uplink response. At the start of cyclic operation, the gNB can use a default bundle length derived from connection-quality measurements. Thereafter, the update trigger in (1) prevents unnecessary bundle changes when the link remains stable while reacting to sustained degradation or rapidly varying reception. Downlink and uplink bundles are adapted separately because the two directions may experience different interference and link budgets.

\begin{figure}[t]
\centering
\includegraphics[width=0.74\columnwidth]{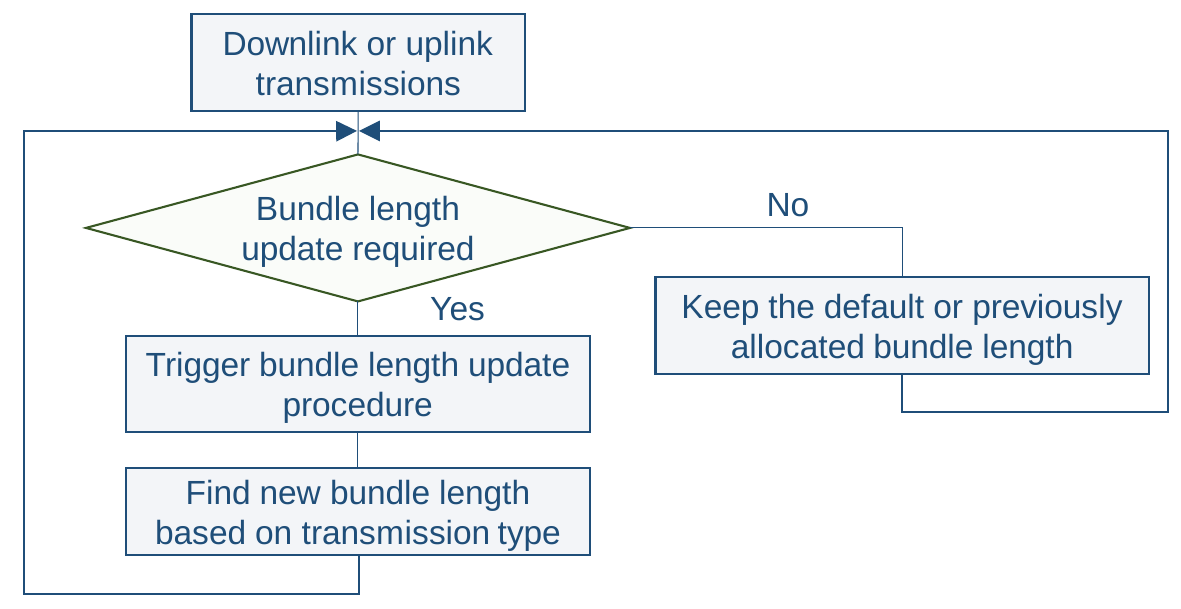}
\caption{Single-user bundle-length update procedure in \Det{}.}
\label{fig:bundle_update}
\end{figure}

For a candidate bundle, let $\Gamma_{\rm single}$ denote the quality of a single transmission and $\Gamma_{\rm gain}(i)$ the reliability/coding gain contributed by the $i$th bundle element. The effective quality is represented as $\Gamma(i)=\Gamma_{\rm single}+\Gamma_{\rm gain}(i)$. Algorithm~\ref{alg:su} selects the smallest bundle that satisfies the receiver target plus an interference margin while respecting $B_{\max}$. This preserves the original \Det{} principle: reliability is adapted before the cycle rather than discovered through reactive retransmission delay.

\begin{algorithm}[t]
\caption{Bundle-length update (single-user)}\label{alg:su}
\KwIn{$\Gamma_{\rm Th}$, $\mathrm{RSSI}_{\rm Avg}$, $I_M$, $B_{\max}$, previous/default $B$}
Set $\mathrm{RSSI}_{\rm single}\leftarrow\mathrm{RSSI}_{\rm Avg}$ and obtain $\Gamma_{\rm single}$\;
Set $\Gamma_{\rm bundle}\leftarrow\Gamma_{\rm single}$ and $B\leftarrow B_{\rm prev}$\;
\While{$\Gamma_{\rm bundle}<\Gamma_{\rm Th}+I_M$ and $B<B_{\max}$}{
  $B\leftarrow B+1$\;
  Update $\Gamma_{\rm bundle}$ using the additional bundle element\;
}
\KwOut{$B$}
\end{algorithm}

\subsection{Multi-user closed-loop control}
\begin{figure}[t]
\centering
\includegraphics[width=0.94\columnwidth]{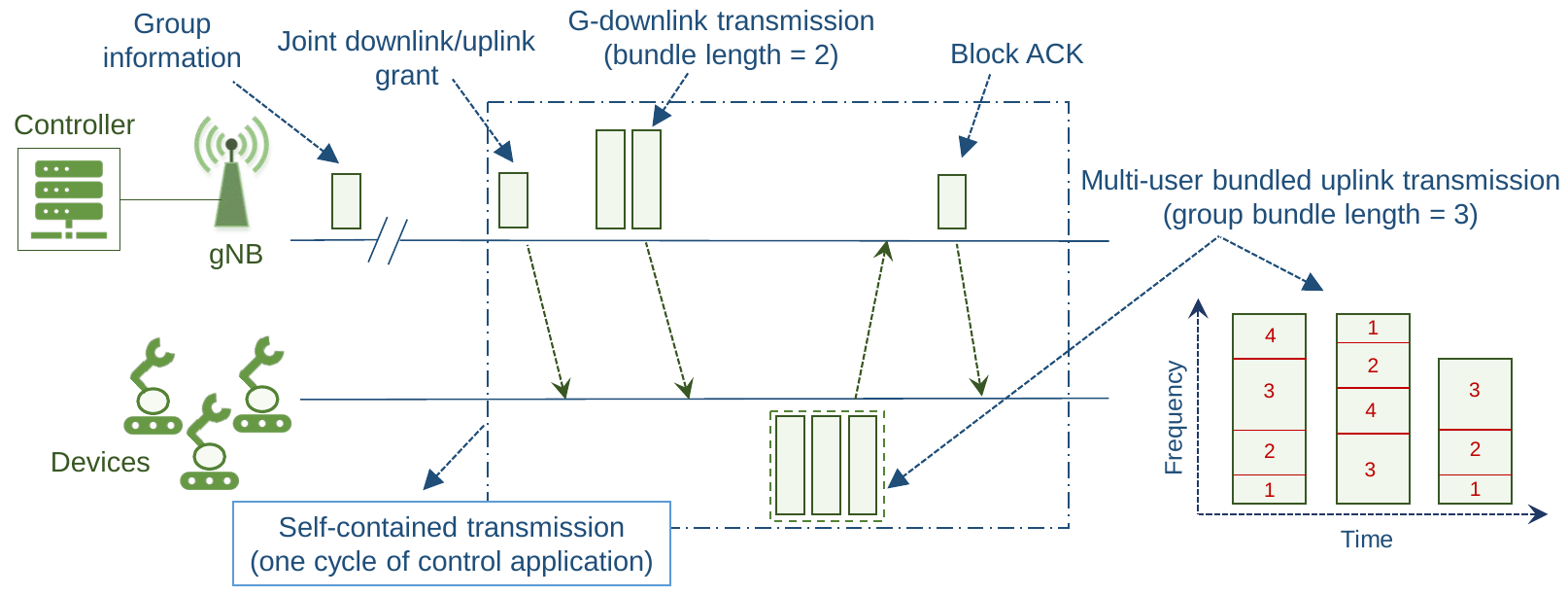}
\caption{Self-contained cyclic transmission for multi-user \Det{}.}
\label{mu_self_cont}
\end{figure}

Directly repeating the single-user procedure for every device would make both command and feedback airtime scale with group size. Multi-user \Det{} therefore modifies the self-contained cycle in two ways: (i) a \emph{group downlink} (G-downlink) delivers the controller command to the complete group and (ii) a \emph{multi-user bundled uplink} packs feedback transmissions from multiple devices into the same time-frequency region. Fig.~\ref{mu_self_cont} shows the resulting cycle. The joint allocation precedes the G-downlink bundle, followed by the packed uplink bundle and a block acknowledgement after successful reception from the group.

A logical group identifier enables the devices belonging to the control group to decode the common downlink transmission. Groups may be predefined in static or quasi-static installations or created dynamically as link conditions, mobility, or control participation change. Dynamic grouping is also useful for recovery: after a self-contained cycle, devices that did not complete the transaction can be regrouped and assigned a subsequent joint allocation instead of forcing successful devices to repeat unnecessarily.

\subsection{G-downlink optimization and group recovery}
Since every member must decode the common command, the G-downlink bundle is governed by the limiting device. For group $\mathcal{M}$, the gNB maintains received-quality statistics for every device and applies the single-user adaptation to the device $w$ with the worst current link. Algorithm~\ref{alg:gdl} therefore turns the individual bundle procedure into a group reliability rule. This worst-device rule guarantees a common downlink reliability target, but it also exposes an important grouping trade-off: repeatedly placing a persistently weak device in a strong group can increase the G-downlink bundle for all members. Dynamic grouping can therefore cluster devices with comparable link conditions or isolate a temporarily degraded device for a subsequent cycle.

The block acknowledgement also operates at group-cycle level. A successful multi-user cycle requires the scheduled uplink response from every participating device; otherwise, the gNB identifies the incomplete subset and can issue a new joint allocation only for that subset. Consequently, successful devices need not repeat their command/feedback exchange solely because another member failed. The same principle can be used with periodic reservation: the regular group schedule remains reserved, while exceptional recovery cycles are inserted for failed members. This keeps recovery aligned with the control transaction rather than with individual packet retransmissions.

\begin{algorithm}[t]
\caption{Bundle-length update for G-downlink}\label{alg:gdl}
\KwIn{Group $\mathcal{M}$ and inputs of Algorithm~\ref{alg:su}}
Find $w\in\mathcal{M}$ with the worst received link quality\;
Run Algorithm~\ref{alg:su} for device $w$\;
Use the resulting $B_{\rm GDL}$ for the G-downlink bundle\;
\end{algorithm}

\subsection{Multi-user bundled uplink optimization}
For the uplink, each device can require a different number of repetitions and a different number of resource blocks (RBs) because its link quality and MCS differ. \Det{} therefore optimizes the bundled uplink in both time and frequency. Two cases are distinguished according to whether the self-contained transmission uses SBTs or NSBTs.

For SBT operation, an RB is the scheduling unit in frequency and the selected SBT interval is the scheduling unit in time. Each device is first mapped from its measured link quality to an MCS and hence to the RB demand needed for its payload, headers, and redundancy. Let $B_m^{\min}$ denote the bundle required by device $m$, $\Gamma_m$ its received quality, and $N_{\rm RB}$ the RB pool in an SBT. The objective is to minimize the group bundle length $G_{\rm BL}$ while meeting every device's bundle and quality requirements:
\begin{equation}
\min G_{\rm BL}\quad
\mathrm{s.t.}\ B_m\geq B_m^{\min},\ \Gamma_m\geq\Gamma_{\rm Th},\ \forall m\in\mathcal{M}.
\label{eq:sbt}
\end{equation}
Algorithm~\ref{alg:sbt} first determines the individual bundle requirements and then packs transmissions into the available RB pool. The group bundle length increases only when another SBT is required.

\begin{algorithm}[t]
\caption{Multi-user uplink optimization (SBTs)}\label{alg:sbt}
\KwIn{$\mathcal{M}$, $\mathcal{N}_{\rm RB}$, MCS table $\mathcal{Q}$ and inputs of Algorithm~\ref{alg:su}}
Find required bundle length $B_m^{\min}$ for every device using Algorithm~\ref{alg:su}\;
Sort devices in descending order of received SNR; set $G_{\rm BL}\leftarrow0$\;
\While{some device has fewer than $B_m^{\min}$ scheduled transmissions}{
  $G_{\rm BL}\leftarrow G_{\rm BL}+1$; reset available RBs\;
  \ForEach{eligible device $m$}{
    Map $\mathrm{SNR}_m$ to $\mathcal{Q}$ and determine its RB demand\;
    Allocate RBs if capacity remains; update the scheduled bundle of $m$\;
  }
}
\KwOut{$G_{\rm BL}$ and the SBT allocation}
\end{algorithm}

For NSBT operation, the allocation is refined to symbol-level time-frequency regions rather than forcing every transmission to occupy the full SBT duration. A device with a high-rate MCS may therefore finish in only a few symbols while a weaker device occupies a longer region. Let $T_n$ denote its required symbol duration. The scheduler minimizes the occupied bundled NSBT length $S_L$ subject to the same reliability/quality constraints. The longest transmission is scheduled first and residual time-frequency opportunities are filled by other devices, as summarized in Algorithm~\ref{alg:nsbt}. This allows short transmissions to use otherwise idle symbols and is particularly effective for heterogeneous links and payloads.

\begin{algorithm}[t]
\caption{Multi-user uplink optimization (NSBTs)}\label{alg:nsbt}
\KwIn{Inputs of Algorithm~\ref{alg:sbt}}
Find $B_n^{\min}$, RB demand, and transmission duration $T_n$ for each device\;
Define $\mathcal{T}_{\rm Txn}$ as the set of outstanding transmissions\;
\While{$\mathcal{T}_{\rm Txn}\neq\emptyset$}{
  Select the outstanding transmission with largest $T_n$\;
  Allocate its required time-frequency resources\;
  Schedule transmissions of other devices in residual feasible opportunities\;
  Update $\mathcal{T}_{\rm Txn}$\;
}
\KwOut{Bundled NSBT allocation}
\end{algorithm}

\section{Performance Evaluation}
The evaluation combines closed-form timing/reliability analysis with a customized 5G air-interface scheduler and Monte Carlo simulations. Figs.~\ref{fig:determinism}, \ref{fig:jitter}, and \ref{fig:scs} are analytical under the stated link/recovery model; hence they have no sampling uncertainty. Fig.~\ref{fig:scale} uses 10,000 independent scheduler realizations per device-count point and reports distribution-free 95\% order-statistic confidence intervals for the 99th percentile. Fig.~\ref{fig:mobility} uses 30 independent runs of 100,000 cycles per speed and reports 95\% confidence intervals.

We use 5G-NR configurations and a periodic industrial motion-control traffic profile with 64-byte messages per direction over a 3.8~GHz\footnote{The 3.8-4.2 GHz band is available for private 5G deployments in the UK.}, 40~MHz carrier. The main configuration uses 30~kHz SCS, a 7-symbol SBT of 0.25~ms, and 106 RBs. CQI/MCS is selected according to current link condition for resource sizing. BLER values $10^{-3}$, $10^{-2}$, and $10^{-1}$ represent favorable, moderate, and challenging states with probabilities 0.50, 0.35, and 0.15. A per-direction residual target $10^{-5}$ gives an approximate cycle target $2\times10^{-5}$. Dynamic and SPS/CG baselines use reactive HARQ; fixed $K=2$ and $K=5$ isolate proactive repetition. The main recovery increment is 1~ms (four SBT quanta). Table~\ref{tab:eval} summarizes the assumptions \cite{5gacia_traffic,3gpp.38.104}.

For the analytical evaluation, the bundle assigned to link state $s$ is the smallest integer whose residual per-direction error probability meets $\epsilon_{\rm dir}$. Under the independent repetition model used for the comparison,
{\setlength{\abovedisplayskip}{3pt}\setlength{\belowdisplayskip}{3pt}
\begin{equation}
B_s=\left\lceil\frac{\ln \epsilon_{\rm dir}}{\ln p_s}\right\rceil,
\label{eq:eval_bundle}
\end{equation}}
where $p_s$ is the BLER of state $s$. This mapping is an evaluation model for quantifying the proactive reliability budget; it is consistent with the design objective of Algorithm~\ref{alg:su}, while the protocol itself may obtain the bundle from measured link quality and coding gain.

For the reactive HARQ baselines, we model the numbers of failed DL and UL transmissions before successful delivery as independent geometric random variables. From their aggregate recovery process, we derive the final expressions used for the cycle deadline-failure probability and expected completion time as
{\setlength{\abovedisplayskip}{3pt}\setlength{\belowdisplayskip}{3pt}\small
\begin{equation}
\begin{aligned}
P_{\rm fail,HARQ}(D)
&=\sum_s\pi_s\!\left[1-\sum_{n=0}^{n_D}(n+1)(1-p_s)^2p_s^n\right],\\
\overline{T}_{\rm HARQ}
&=\sum_s\pi_s\!\left(T_0+\frac{2p_s}{1-p_s}T_R\right).
\end{aligned}
\label{eq:harq_model}
\end{equation}}
Here $n_D=\lfloor(D-T_0)/T_R\rfloor$ is the number of recovery opportunities that fit before deadline $D$, $T_0$ is the no-recovery cycle time, $T_R$ is the recovery interval, $p_s$ is the BLER of state $s$, and $\pi_s$ is its occurrence probability. For the single-user \Det{} comparison, the corresponding average is $\overline{T}_{\rm Det}=\sum_s\pi_s(C+2B_s)T_q$, where $C=3$ fixed allocation/control quanta and $T_q$ is the transmission-quantum duration.

\begin{table}[t]
\caption{Evaluation Parameters}
\label{tab:eval}
\centering
\footnotesize
\begin{tabular}{@{}ll@{}}
\toprule
Parameter & Setting \\
\midrule
Carrier / bandwidth & 3.8~GHz / 40~MHz \\
Message size & 64~bytes per direction \\
Main SCS / SBT & 30~kHz / 7 symbols (0.25~ms) \\
Available RBs & 106 \\
Link state probabilities & 0.50 / 0.35 / 0.15 \\
BLER values & $10^{-3}/10^{-2}/10^{-1}$ \\
MCS selection & CQI-driven, link-adaptive \\
Main deadline / cycle target & 4~ms / $2\times10^{-5}$ \\
Stringent stress point (Fig.~\ref{fig:scs}) & 1~ms / $10^{-6}$ \\
Recovery interval & 4 transmission quanta \\
Controlled devices / mobility & 1--20 / 0--10~m/s \\
\bottomrule
\end{tabular}
\end{table}

\subsection{Joint cycle time and deadline reliability}
\begin{figure}[t]
\centering
\includegraphics[width=0.90\columnwidth]{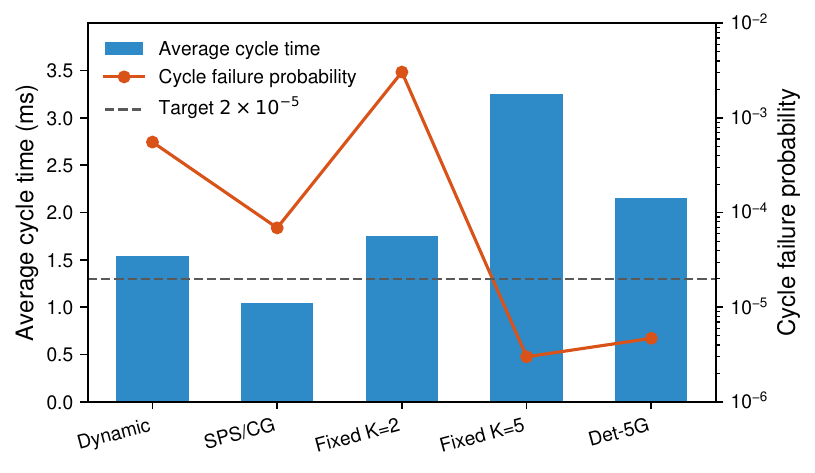}
\caption{Average cycle time and cycle failure probability under a common 4~ms deadline. Values are obtained analytically; the horizontal line is the cycle-failure target.}
\label{fig:determinism}
\end{figure}

Fig.~\ref{fig:determinism} evaluates the central trade-off under the common 4~ms application deadline. The left axis is the exact expected eventual cycle completion time under the link-state mixture; the right axis is the probability that the complete downlink/uplink transaction misses the deadline or is not delivered successfully. For reactive HARQ, the aggregate number of DL and UL failures follows a negative-binomial distribution. For fixed repetition, the residual cycle failure at BLER $p$ is $1-(1-p^K)^2$; \Det{} uses the corresponding state-dependent bundle $B$.

Dynamic scheduling averages 1.54~ms but has cycle failure $5.6\times10^{-4}$; SPS/CG averages 1.04~ms but remains at $6.9\times10^{-5}$. Fixed $K=2$ is short (1.75~ms) but unreliable at $3.1\times10^{-3}$. Fixed $K=5$ reaches $3.0\times10^{-6}$ but fixes every cycle at 3.25~ms. \Det{} reaches $4.7\times10^{-6}$ with 2.15~ms average cycle time. Thus average latency alone is insufficient: relative to fixed $K=5$, \Det{} lowers average cycle time by about 34\% and average proactive repetition occupancy by 44\% while satisfying the target.

\subsection{Retransmission-induced cycle time variation}
\begin{figure}[t]
\centering
\includegraphics[width=0.88\columnwidth]{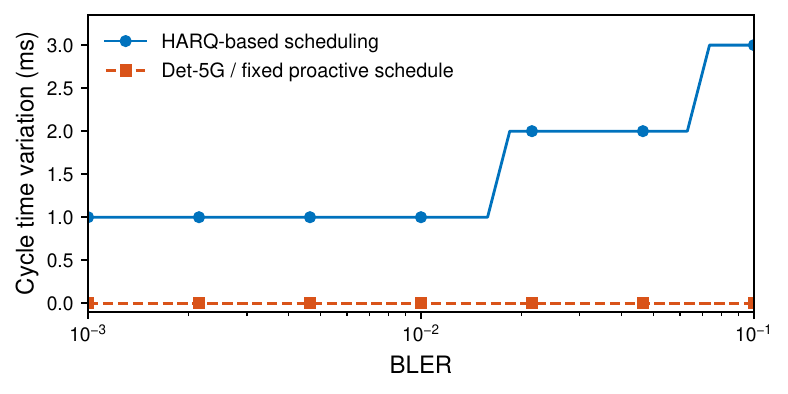}
\caption{Analytical retransmission-induced cycle time variation, measured as the 99.9th--0.1th percentile spread at fixed BLER.}
\label{fig:jitter}
\end{figure}

Fig.~\ref{fig:jitter} isolates reactive timing uncertainty at fixed BLER. The 99.9th--0.1th percentile spread is calculated directly from the exact negative-binomial distribution of aggregate DL/UL HARQ failures, rather than estimated from finite trials. HARQ-based scheduling develops 1--3~ms of cycle time variation across the evaluated BLER range. \Det{} and fixed proactive schedules have zero \emph{reactive} variation once a cycle allocation is selected; link adaptation may change the predetermined budget between cycles but does not append an unplanned HARQ tail to the allocated cycle.

\subsection{Multi-user cyclic scalability}
\begin{figure}[t]
\centering
\includegraphics[width=0.89\columnwidth]{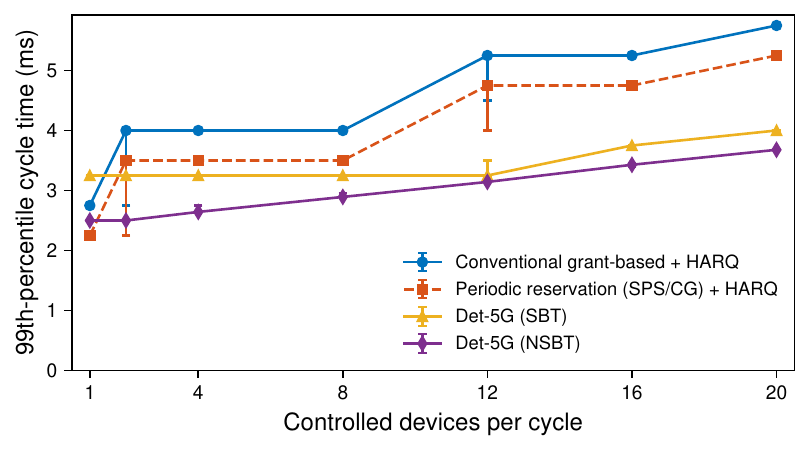}
\caption{99th-percentile cycle time with multi-user frequency multiplexing enabled for all schemes. Error bars are 95\% order-statistic confidence intervals.}
\label{fig:scale}
\end{figure}

Fig.~\ref{fig:scale} removes the weak sequential baseline used in an earlier formulation. The conventional dynamic and SPS/CG schedulers are allowed to frequency-multiplex independent UE transmissions within the same SBT using the same 106-RB pool and CQI-dependent RB demand as \Det{}. They retain per-UE DL/UL transactions and reactive HARQ, whereas \Det{} uses one G-downlink and proactively packs bundled uplink transmissions. The figure therefore compares the 99th-percentile cycle time rather than mean latency, since the latter favors reactive schemes when no recovery is needed. Resource demand is computed consistently for all schemes as $N_{\rm RB}=\lceil L/(\eta_{\rm CQI}N_{\rm RE})\rceil$, where $L$ is the transmitted bit count, $\eta_{\rm CQI}$ is the spectral efficiency associated with the sampled CQI, and $N_{\rm RE}$ is the usable resource-element count for the selected transmission duration. In the NSBT model, CQI values 11--15, 7--10, and 1--6 use 2-, 4-, and 7-symbol transmission durations, respectively.

The fairer comparison substantially reduces the previously observed gain and changes the interpretation. At one or two devices the conventional baselines have lower tail cycle time. As group size grows, however, the probability that at least one independent transaction requires recovery increases. At 16 devices, the 99th-percentile cycle times are 5.25~ms for dynamic scheduling, 4.75~ms for SPS/CG, 3.75~ms for \Det{} SBT, and 3.43~ms for \Det{} NSBT. Thus NSBT lowers the 99th-percentile cycle time by about 35\% and 28\% relative to the dynamic and SPS/CG baselines, respectively. Error bars show 95\% order-statistic confidence intervals from 10,000 independent realizations per point.

\subsection{Stringent latency/reliability stress test}
\begin{figure}[t]
\centering
\includegraphics[width=0.82\columnwidth]{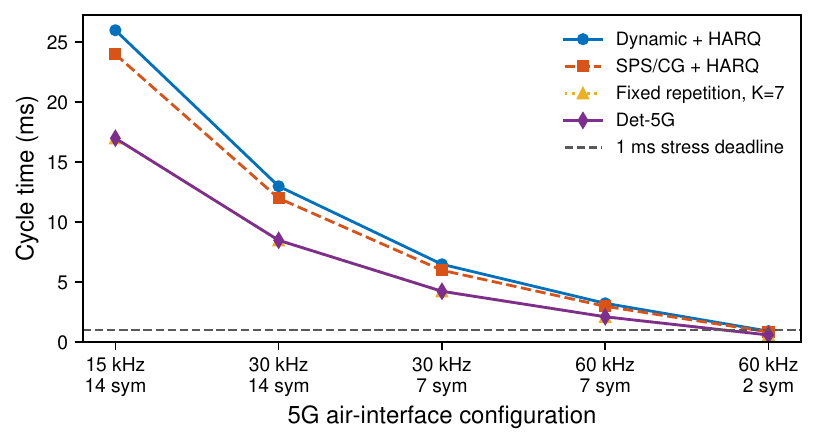}
\caption{Analytical cycle time across 5G air-interface configurations at a $10^{-6}$ cycle-failure target. The 1~ms line is a stringent full-cycle stress deadline; fixed $K=7$ coincides with the \Det{} worst-state timing bound.}
\label{fig:scs}
\end{figure}

The 5G-ACIA motion-control SLS example lists 1~ms end-to-end latency, 1~ms transfer interval, zero survival time, and 99.9999\% communication-service availability \cite{5gacia_sls}. Availability is a service-level quantity and is not identical to the per-cycle radio failure probability used here, so we do not convert one directly into the other. Instead, Fig.~\ref{fig:scs} adds an intentionally stringent radio-side stress point: a 1~ms full-cycle deadline and $10^{-6}$ cycle-failure target.

Meeting $10^{-6}$ requires five aggregate HARQ recovery opportunities for the reactive baselines under the link-state mixture. \Det{} requires state-dependent bundles $B=3,4,7$; fixed $K=7$ therefore has the same worst-state timing bound but repeats seven times in every state. At 60~kHz/7 symbols, the resulting cycle times remain 3.25~ms (dynamic), 3.0~ms (SPS/CG), and 2.125~ms (\Det{}/fixed $K=7$). With a 60~kHz 2-symbol NSBT opportunity they fall to 0.93, 0.86, and 0.61~ms, respectively. Hence the strict 1~ms full-cycle stress point is reached only with sufficiently fine transmission granularity in the evaluated set; \Det{} provides the largest timing margin while avoiding persistent worst-case repetition.

\subsection{Mobility and bundle update overhead}
\begin{figure}[t]
\centering
\includegraphics[width=0.78\columnwidth]{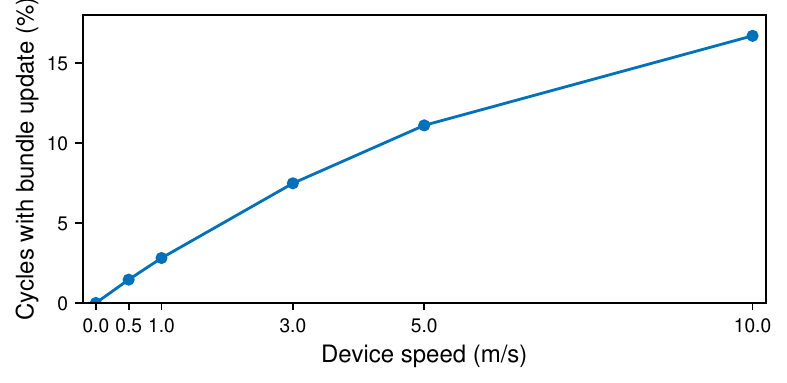}
\caption{Event-triggered \Det{} bundle update frequency versus device speed. }
\label{fig:mobility}
\end{figure}

Mobility does not imply that \Det{} must change its bundle every control cycle. Link quality is measured continuously, while the bundle changes only after a persistent state transition. We use $T_c\approx0.423/f_D$ at 3.8~GHz and a two-cycle persistence rule. The coherence time is mapped to a per-cycle state-transition probability $q=1-\exp(-T_{\rm cp}/T_c)$, where $T_{\rm cp}=1$~ms is the reference control period used in this sensitivity study. Fig.~\ref{fig:mobility} reports the mean update frequency from 30 independent runs of 100,000 cycles per speed. Bundle updates occur in about 2.81\%, 7.48\%, 11.10\%, and 16.70\% of cycles at 1, 3, 5, and 10~m/s; all nonzero-speed 95\% confidence half-widths are below 0.04 percentage points.

Relative to cycle-by-cycle \Det{}, event-triggered reserved operation therefore reduces bundle update actions by approximately 97.2\%, 92.5\%, 88.9\%, and 83.3\% at 1, 3, 5, and 10~m/s. SPS/CG can require less routine reconfiguration, but recovery still introduces timing uncertainty; fixed worst-case repetition requires little adaptation but pays continuous resource cost. Reserved \Det{} occupies the middle ground: reliability resources change only when the link state persistently changes, while the allocated cycle remains predetermined.

\section{Concluding Remarks}

This paper introduced \Det{}, a cycle-oriented radio resource allocation solution for deterministic industrial closed-loop control over 5G.  \Det{} treats the complete bidirectional control cycle as the scheduling unit and combines coordinated downlink/uplink allocation, proactive link-adaptive bundled transmissions, cycle-level recovery, group-oriented downlink communication, and SBT/NSBT-based multi-user uplink optimization. The resulting design aims to bound control-cycle completion while maintaining reliability and adapting radio resource use to changing link conditions and multi-device operation. Under the main 4~ms operating point, dynamic and SPS/CG scheduling achieve shorter average cycle times but exceed the cycle-failure target, whereas \Det{} satisfies the target with a 2.15~ms average cycle and uses 44\% less proactive repetition occupancy than fixed $K=5$. Analytical characterization further shows that reactive HARQ introduces 1--3~ms cycle time variation. With multi-user frequency multiplexing enabled for the conventional baselines, \Det{} NSBT provides a 28--35\% reduction in 99th-percentile cycle time at 16 devices. The stringent stress study further shows that a 1~ms full-cycle/$10^{-6}$ radio target requires sufficiently fine transmission granularity, with \Det{} providing the largest timing margin among the evaluated schemes. Mobility adaptation remains event-triggered, requiring bundle updates in only about 7.5\% of cycles at 3~m/s.

These results position \Det{} as a system-level way to address an unresolved industrial 5G adoption barrier as networks evolve through 5G-Advanced, while using existing NR mechanisms rather than depending on a new 5G-Advanced primitive. The framework is also complementary to 5G/TSN convergence: synchronization, QoS coordination, and deterministic Ethernet integration do not remove radio-side timing uncertainty. Bounding the complete wireless control transaction can therefore help extend deterministic behavior across converged wired/wireless systems and support progression from connectivity-oriented private 5G toward closed-loop automation, coordinated robotics, Physical AI, and ultimately 6G cyber-physical communication.

\bibliographystyle{IEEEtran}
\balance
\bibliography{bibliography_final_v2}

\begin{thebibliography}{10}
\providecommand{\url}[1]{#1}
\csname url@samestyle\endcsname
\providecommand{\newblock}{\relax}
\providecommand{\bibinfo}[2]{#2}
\providecommand{\BIBentrySTDinterwordspacing}{\spaceskip=0pt\relax}
\providecommand{\BIBentryALTinterwordstretchfactor}{4}
\providecommand{\BIBentryALTinterwordspacing}{\spaceskip=\fontdimen2\font plus
\BIBentryALTinterwordstretchfactor\fontdimen3\font minus
  \fontdimen4\font\relax}
\providecommand{\BIBforeignlanguage}[2]{{%
\expandafter\ifx\csname l@#1\endcsname\relax
\typeout{** WARNING: IEEEtran.bst: No hyphenation pattern has been}%
\typeout{** loaded for the language `#1'. Using the pattern for}%
\typeout{** the default language instead.}%
\else
\language=\csname l@#1\endcsname
\fi
#2}}
\providecommand{\BIBdecl}{\relax}
\BIBdecl

\bibitem{pvt_5G}
A.~{Aijaz}, ``{Private 5G: The Future of Industrial Wireless},'' \emph{IEEE
  Ind. Electron. Mag.}, vol.~14, no.~4, pp. 136--145, 2020.

\bibitem{3gpp_tsn}
\BIBentryALTinterwordspacing
{3GPP}, ``{5G for Industry 4.0},'' 2020, overview of Release-16 TSN/TSC, NPN
  and 5G-LAN support. [Online]. Available:
  \url{https://www.3gpp.org/technologies/tsn-v-lan}
\BIBentrySTDinterwordspacing

\bibitem{3gpp_22104}
------, ``{Service Requirements for Cyber-Physical Control Applications in
  Vertical Domains},'' 3rd Generation Partnership Project (3GPP), TS 22.104,
  2026, specification under change control.

\bibitem{5gacia_sls}
\BIBentryALTinterwordspacing
{5G Alliance for Connected Industries and Automation (5G-ACIA)},
  ``{Service-Level Specifications (SLSs) for 5G Technology-Enabled Connected
  Industries},'' 5G-ACIA / ZVEI, Tech. Rep., Sep. 2021, motion-control example:
  1 ms end-to-end latency, 1 ms transfer interval, zero survival time, and
  99.9999 percent communication-service availability. [Online]. Available:
  \url{https://5g-acia.org/download/18765/?version=A4}
\BIBentrySTDinterwordspacing

\bibitem{5gacia_test}
\BIBentryALTinterwordspacing
------, ``{Performance Testing of 5G Systems for Industrial Automation},''
  2021, white paper. [Online]. Available:
  \url{https://5g-acia.org/whitepapers/performance-testing-of-5g-systems-for-industrial-automation/}
\BIBentrySTDinterwordspacing

\bibitem{3gpp_rel20}
\BIBentryALTinterwordspacing
{3GPP}, ``{Release 20},'' 2026, 5G-Advanced technical specifications and early
  6G studies. [Online]. Available:
  \url{https://www.3gpp.org/specifications-technologies/releases/release-20}
\BIBentrySTDinterwordspacing

\bibitem{li_networked_robotics_6g}
P.~Li, X.~Lin, and A.~Aijaz, ``{Rethinking Networked Robotics in the 6G Era
  With Generative AI-in-the-Loop},'' \emph{IEEE Communications Standards
  Magazine}, vol.~10, no.~2, pp. 287--295, 2026.

\bibitem{enclose}
A.~Aijaz, ``{ENCLOSE: An Enhanced Wireless Interface for Communication in
  Factory Automation Networks},'' \emph{IEEE Trans. Ind. Informat.}, vol.~14,
  no.~12, pp. 5346--5358, 2018.

\bibitem{gallop}
A.~Aijaz and A.~Stanoev, ``{Closing the Loop: A High-Performance Connectivity
  Solution for Realizing Wireless Closed-Loop Control in Industrial IoT
  Applications},'' \emph{IEEE Internet Things J.}, vol.~8, no.~15, pp.
  11\,860--11\,876, 2021.

\bibitem{5G_FA}
S.~A. {Ashraf} \emph{et~al.}, ``{Ultra-reliable and Low-latency Communication
  for Wireless Factory Automation: From LTE to 5G},'' in \emph{IEEE ETFA},
  2016, pp. 1--8.

\bibitem{retrans_URLLC}
S.~E. {Elayoubi}, P.~{Brown}, M.~{Deghel}, and A.~{Galindo-Serrano}, ``{Radio
  Resource Allocation and Retransmission Schemes for uRLLC Over 5G Networks},''
  \emph{IEEE J. Sel. Areas Commun.}, vol.~37, no.~4, pp. 896--904, 2019.

\bibitem{URLLC_control1}
B.~{Chang} \emph{et~al.}, ``{Optimizing Resource Allocation in uRLLC for
  Real-Time Wireless Control Systems},'' \emph{IEEE Trans. Veh.Tech.}, vol.~68,
  no.~9, pp. 8916--8927, 2019.

\bibitem{demel2020burst}
J.~S. Demel, C.~Bockelmann, and A.~Dekorsy, ``{Burst Error Analysis of
  Scheduling Algorithms for 5G NR uRLLC Periodic Deterministic
  Communication},'' in \emph{2020 IEEE 91st Vehicular Technology Conference
  (VTC2020-Spring)}, 2020, pp. 1--5.

\bibitem{sch_TSN1}
N.~Jiang, A.~Aijaz, and Y.~Jin, ``{Recursive Periodicity Shifting for
  Semi-Persistent Scheduling of Time-Sensitive Communication in 5G},'' in
  \emph{IEEE Global Communications Conference (GLOBECOM)}, 2021, pp. 01--06.

\bibitem{zhang2023cg}
T.~Zhang, X.~S. Hu, and S.~Han, ``{Contention-Free Configured Grant Scheduling
  for 5G uRLLC Traffic},'' in \emph{ACM/IEEE Design Automation Conference
  (DAC)}, 2023, pp. 1--6.

\bibitem{zhang2023rtss}
T.~Zhang, J.~Wang, X.~S. Hu, and S.~Han, ``{Real-Time Flow Scheduling in
  Industrial 5G New Radio},'' in \emph{IEEE Real-Time Systems Symposium
  (RTSS)}, 2023, pp. 371--384.

\bibitem{kleinberger2026flex}
L.~Kleinberger, M.~Gundall, and H.~D. Schotten, ``{FLEX: Joint UL/DL and
  QoS-Aware Scheduling for Dynamic TDD in Industrial 5G and Beyond},'' in
  \emph{IEEE International Conference on Communications (ICC)}, 2026, preprint
  arXiv:2603.20971.

\bibitem{3gpp.38.214}
3GPP, ``{NR; Physical Layer Procedures for Data},'' {3rd Generation Partnership
  Project (3GPP)}, TS {38.214}, Jun. 2026, {Release 19, v19.4.0}.

\bibitem{5gacia_traffic}
\BIBentryALTinterwordspacing
{5G Alliance for Connected Industries and Automation (5G-ACIA)}, ``{A 5G
  Traffic Model for Industrial Use Cases},'' 2019, white paper; motion-control
  traffic includes small periodic messages and 1 ms sensor transfer intervals.
  [Online]. Available:
  \url{https://5g-acia.org/whitepapers/a-5g-traffic-model-for-industrial-use-cases/}
\BIBentrySTDinterwordspacing

\bibitem{3gpp.38.104}
{3GPP}, ``{NR; Base Station (BS) Radio Transmission and Reception},'' 3rd
  Generation Partnership Project (3GPP), TS 38.104, 2026, release 19.

\end{thebibliography}
\end{document}